\documentclass[conference]{IEEEtran}
\IEEEoverridecommandlockouts
\usepackage{cite}
\usepackage{soul}
\usepackage{xcolor}
\sethlcolor{yellow}

\usepackage{amsmath,amssymb,amsfonts}
\usepackage{algorithmic}
\usepackage{graphicx}
\usepackage{textcomp}
\usepackage{xcolor}
\usepackage{booktabs}
\usepackage{url}
\def\BibTeX{{\rm B\kern-.05em{\sc i\kern-.025em b}\kern-.08em
    T\kern-.1667em\lower.7ex\hbox{E}\kern-.125emX}}
    
\begin{document}

\title{Let's have a Chat with the EU AI Act\\
\thanks{This work is done within the AIMS5.0 project which is supported by the Chips JU and its members, including the top-up funding from national funding authorities from involved countries under grant agreement no. 101112089.}
}

\author{\IEEEauthorblockN{%
\fontsize{13}{16}\selectfont
Ádám Kővári \IEEEauthorrefmark{2},
Yasin Ghafourian \IEEEauthorrefmark{1},
Csaba Hegedűs\IEEEauthorrefmark{2},
Belal Abu Naim\IEEEauthorrefmark{1},
Kitti Mezei \IEEEauthorrefmark{2}\IEEEauthorrefmark{3},\\
Pál Varga \IEEEauthorrefmark{2},
Markus Tauber \IEEEauthorrefmark{1},
}
\vspace{+10pt}

\IEEEauthorblockA{\IEEEauthorrefmark{1}%
\fontsize{12}{16}\selectfont 
Research Studios Austria,
\IEEEauthorrefmark{2}%
Budapest University of Technology and Economics,\\
and \IEEEauthorrefmark{3}%
HUN-REN Centre for Social Sciences
}

}

\maketitle

\begin{abstract}
As artificial intelligence (AI) regulations evolve and the regulatory landscape develops and becomes be more complex, ensuring compliance with ethical guidelines and legal frameworks remains a challenge for AI developers. This paper introduces an AI-driven self-assessment chatbot designed to assist users in navigating the European Union AI Act and related standards. Leveraging a Retrieval-Augmented Generation (RAG) framework, the chatbot enables real-time, context-aware compliance verification by retrieving relevant regulatory texts and providing tailored guidance. By integrating both public and proprietary standards, it streamlines regulatory adherence, reduces complexity, and fosters responsible AI development. The paper explores the chatbot’s architecture, comparing naive and graph-based RAG models, and discusses its potential impact on AI governance.
\end{abstract}


\section{Introduction}
The rapid evolution of artificial intelligence (AI) technologies has enabled transformative applications across industries that are empowered by AI components and services. Nevertheless, it has also introduced significant challenges related to ethics, legal compliance, and governance. As global and regional regulations, such as the European Union’s AI Act~\cite{eu_ai_act}, gain traction, developers and publishers of industrial AI systems face increasing complexity in navigating diverse guidelines, standards, and legal frameworks. This complexity is compounded by the need to ensure ethical soundness, regulatory adherence, and conformance to national and international standards. Although there are tools based on Large Language Models (LLMs) that help the users to navigate these documents, such as ChatGPT, or tools provided by standardization organizations, these tools are either fundamental inquiry tools or very generic and do not cover copyrighted standards.\looseness=-1

To address this challenge, we are building an innovative AI-driven self-assessment tool—a chatbot designed to assist AI developers and users in evaluating the compliance of their applications with ethical guidelines, legal regulations, and industry standards. Leveraging a Retrieval-Augmented Generation (RAG) framework, and hosted on cloud infrastructure, the chatbot provides real-time, context-aware guidance, allowing users to assess their AI systems efficiently. By facilitating the compliance verification process, this solution not only encourages a consistent approach to ethics and compliance but also reduces the time and expertise required to navigate complex regulatory landscapes. Furthermore, the chatbot enables developers and AI project managers to access not only public regulations but also company-owned standards and guidelines—whether these are proprietary regulations developed internally or industry standards the organization has acquired. The integration of these standards will allow users to browse through hundreds of regulations more easily and allows for organizations to maintain the recency of their standards by focusing the chatbot's local repository on the most up-to-date versions. By collecting all necessary documents in one system, the chatbot empowers developers and organizations to build AI systems that are not only innovative but also responsible and legally sound.\looseness=-1

The users of this self-assessment tool will be able to interact with the chatbot and identify the relevant and applicable sections from regulatory frameworks that pertain to their AI-empowered applications. By engaging in a conversational interface, users can pose specific questions about their system’s design, deployment, or use-case scenarios and receive tailored, easy-to-understand guidance drawn from a comprehensive repository of legal and ethical guidelines. This is enabled by the descriptive power of LLMs, which allow the chatbot to interpret user queries in nuanced ways, extract precise information from complex documents, and present it in a user-friendly manner.\looseness=-1

This paper introduces the architecture, functionalities, and implications of our chatbot which is designed as a self-assessment tool for AI ethics and compliance. To the best of our knowledge, it is the first attempt to utilize LLMs and RAG to create a tool that navigates the regulatory landscape and facilitates compliance for the AI community. Additionally, we highlight key ethical considerations, user-centric design principles, and the tool’s potential to foster responsible AI development.\looseness=-1  

Section~2 will provide an overview of the complex and evolving regulatory landscape surrounding AI.
In Section~3 we will discuss the architecture of our self-assessment tool.
Section~4 will conclude the paper and will discuss the future plans towards extending the current self-assessment tool.\looseness=-1 


\section{The Regulatory Landscape}
The regulatory landscape surrounding AI is rapidly evolving as governments and organizations worldwide work to address the ethical, legal, and societal challenges posed by AI technologies. This landscape is diverse and multifaceted, encompassing a range of document categories that vary in scope, nature, and focus areas. These categories include high-level frameworks and guidelines, such as the OECD AI Principles \cite{oecd_ai_principles} and UNESCO’s recommendations \cite{unesco_ai_ethics}, which provide non-binding ethical and governance principles. Binding legislation and statutory laws, like the EU’s AI Act \cite{eu_ai_act} and the General Data Protection Regulation (GDPR)\cite{gdpr}, the Digital Markets Act (DMA)\cite{digital_markets_act}, the Digital Services Act \cite{digital_services_act}, and the Data Governance Act (DGA)\cite{data_governance_act}, establish enforceable rights and obligations, focusing on areas such as data privacy, risk management, and accountability \cite{novelli2024generative}.\looseness=-1 

Technical standards, including those developed by ISO/IEC and IEEE, define operational specifications for AI systems, ensuring performance, interoperability, and safety\cite{cantero2024artificial}. Sector-specific regulations, such as FDA guidelines for AI in medical devices \cite{aboy2024navigating} or autonomous vehicle safety standards or the Cyber Resilience Act (CRA) \cite{cyber_resilience_act}, address unique challenges in critical domains. Additionally, national AI strategies (e.g., Hungary's AI strategy\footnote{https://bitly.cx/HD59x}) and policies set long-term goals for innovation and international competitiveness, while voluntary codes and best practices, often developed by industry leaders, foster ethical development without legal mandates.\looseness=-1 

With regulatory documents spanning ethical guidelines, binding legislation, technical standards, and sector-specific rules—each varying by jurisdiction, focus area, and binding nature—identifying the most relevant requirements can be a daunting task. The self-assessment tool proposed in this paper addresses this challenge by enabling stakeholders to efficiently map their AI applications to the applicable regulations and standards. This not only saves time but also ensures comprehensive compliance, fostering responsible AI development while adapting to the dynamic nature of the regulatory ecosystem. By bridging the gap between complex regulatory frameworks and practical implementation, the self-assessment tool aims to become an essential asset in aligning AI innovations with ethical, legal, and technical benchmarks.\looseness=-1

In their highly related work, Huang et al. \cite{Huang.2023} explored the ethical challenges arising from AI's rapid integration into areas like healthcare, finance, and governance. They highlight key concerns at individual (privacy, safety), societal (bias, accountability, democracy), and environmental levels (resource use, pollution). Reviewing 146 global guidelines, the authors identify core ethical principles: transparency, fairness, responsibility, non-maleficence, and privacy. They discuss approaches to mitigate these challenges, including technological solutions (e.g., explainable AI, fair algorithms), ethical frameworks (virtue ethics, deontology), and legal regulations (e.g., GDPR). Methods for evaluating AI ethics, such as the Moral Turing Test and international standards, are also examined. Their work stresses the need for global consensus, interdisciplinary efforts, and integrated ethical frameworks to ensure AI benefits society while mitigating harm.\looseness=-1

Initially, Abu Naim et al.~\cite{abunaim_10575125} proposed a self-assessment tool to facilitate the adoption of AI in industrial digital workspaces by addressing trust, ethical, and regulatory challenges. They outlined a four-step methodology: (1) defining personas based on expert interviews to categorize AI adopters into early adopters, followers, and independent learners; (2) mapping relevant AI regulations and ethical standards to ensure compliance; (3) fine-tuning LLMs to provide tailored guidance on AI implementation; and (4) developing an AI-powered self-assessment tool integrating profiling, memory, planning, and action modules. The tool aims to bridge the gap between AI innovation and legal compliance, enhancing user trust and promoting sustainable AI adoption in alignment with Industry~5.0 goals.\looseness=-1

\subsection{About the AI Act}
The EU AI Act~\cite{eu_ai_act} is a pioneering regulatory framework that categorizes AI systems by risk level—ranging from minimal to unacceptable—and imposes corresponding obligations. High-risk AI, such as in healthcare or law enforcement, faces strict requirements for transparency, accountability, and data governance. Additionally, the EU AI Act is notable for its emphasis on fostering trust in AI technologies while maintaining a balance between innovation and regulation. By mandating practices like human oversight, bias mitigation, and robust documentation, the AI Act seeks to ensure AI systems operate ethically and safely. Beyond risk mitigation, the Act fosters trust in AI through mandates on human oversight, bias mitigation, and documentation. Its extraterritorial scope, akin to the GDPR, extends to entities outside the EU impacting EU citizens. Rooted in European values of fundamental rights and fairness, the Act sets a global precedent for ethical AI governance \cite{de2022european}. \looseness = -1


\section{System Design and Architecture}
Modern, widely adopted LLMs like GPT-4~\cite{OpenAI_ChatGPT} are well-suited for general-purpose use and text comprehension, as they are trained on an extensive and broad knowledge base. However, this is often not sufficient when accurate and up-to-date information is required. Modern language models often grapple with challenges like limited access to the (i) latest information, (ii) insufficient expertise in specific domains or tasks, and (iii) a tendency to generate inaccurate or fabricated responses due to their probabilistic design.\looseness=-1

The architecture we have designed seeks to address or mitigate these issues through the implementation of a Retrieval-Augmented Generation (RAG) framework. This approach ensures that, during the process of answer generation, the LLM is provided with (i) up-to-date and (ii) domain-specific information relevant to the query. By doing so, it significantly reduces (iii) the likelihood of hallucinations and enhances the accuracy of the generated responses.\looseness=-1

\subsection{Architecture}
In general, the implementation of a RAG system consists of two components. In a preparatory phase, referred to as Indexing, we digitize the regulatory documents (like the EU AI Act). During this phase, information is extracted from the documents, and a database is constructed. Later, in the RAG phase, elements from this database that resemble the information implied by the user's intent are retrieved and provided to the context window of the LLM. The LLM then interprets this information and generates an appropriate response.\looseness=-1

Various types of RAG models are being developed, each exhibiting varying levels of performance depending on the nature of the data and the query derived from it. Initially, we implemented the most basic form, referred to as the Naive RAG~\cite{rag-overview}, to explore the capabilities of a basic approach to the problem and in order to have a benchmark for comparing the results of other RAG types.
In the next phase of the project, we will implement a more structured, hierarchical model called Graph RAG~\cite{Microsoft_GraphRAG}, to assess whether it can show a significant improvement compared to the naive RAG.

\subsection{Naive RAG}

\begin{figure}[!htb] 
	\centering
	\includegraphics[width=0.5\textwidth]{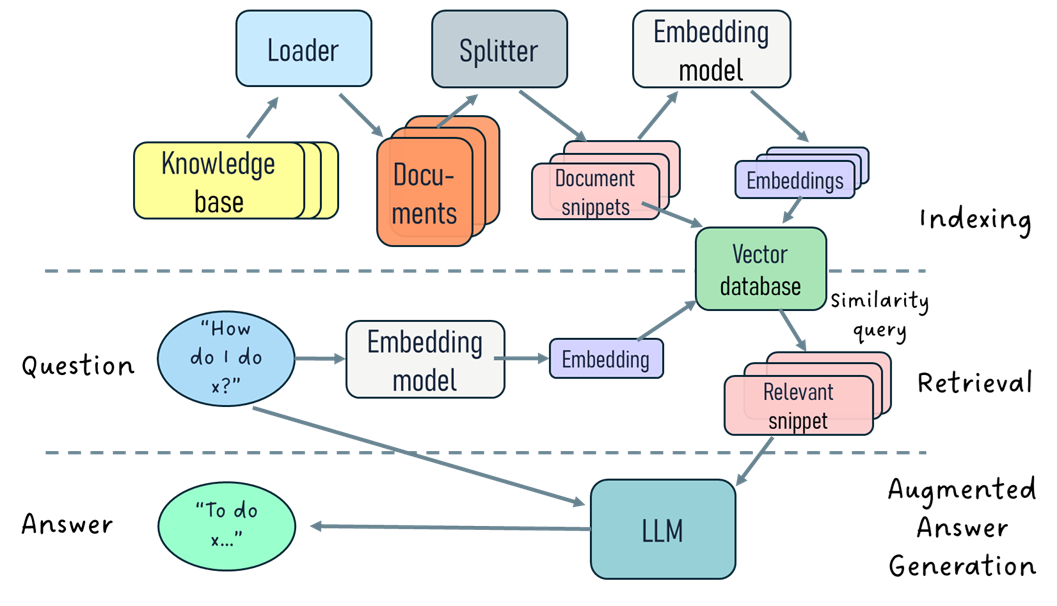}
	\caption{Naive RAG}
	\label{fig:naive-rag}
\end{figure}

\subsubsection{Indexing}
The input to the indexing phase is a document-like file. Using various document processing techniques, like Optical Character Recognition (OCR), we extract text, tables, and visual content, making it possible to perform multimodal document processing. This approach contrasts with traditional indexing methods that only handle textual content, ensuring that valuable information derived from table structures and visual elements is preserved. The extracted information elements are then summarized and transformed into vectors using an embedding model and stored to perform semantic similarity search during retrieval. Fig.~\ref{fig:naive-rag} illustrates the entire process in the upper Indexing section.\looseness=-1

\subsubsection{Retrieval-Augmented Generation}

In the context of RAG, the user submits a question or provides an instruction in textual form. This input is similarly converted into a vector using the same embedding model utilized during the indexing phase. A similarity search is then performed against the vector database containing vector representations of the indexed document segments. The retrieved document segments are subsequently provided to the LLM, along with the user's intent, to facilitate response generation.\looseness=-1

\subsubsection{Functionality}
The system implementing the naive RAG that we developed:

\begin{itemize}
  \item Is capable of processing and integrating multimodal documents into the knowledge base regardless of their length, based on the Table of Contents of the documents.
  \item During chat, it interprets the user intent, searches for similarities in the knowledge base based on semantic analysis, and formulates a response by interpreting the resulting document segments, including source attribution.    
\end{itemize} 
\begin{figure}[!t] 
    \centering
    \includegraphics[width=0.5\textwidth]{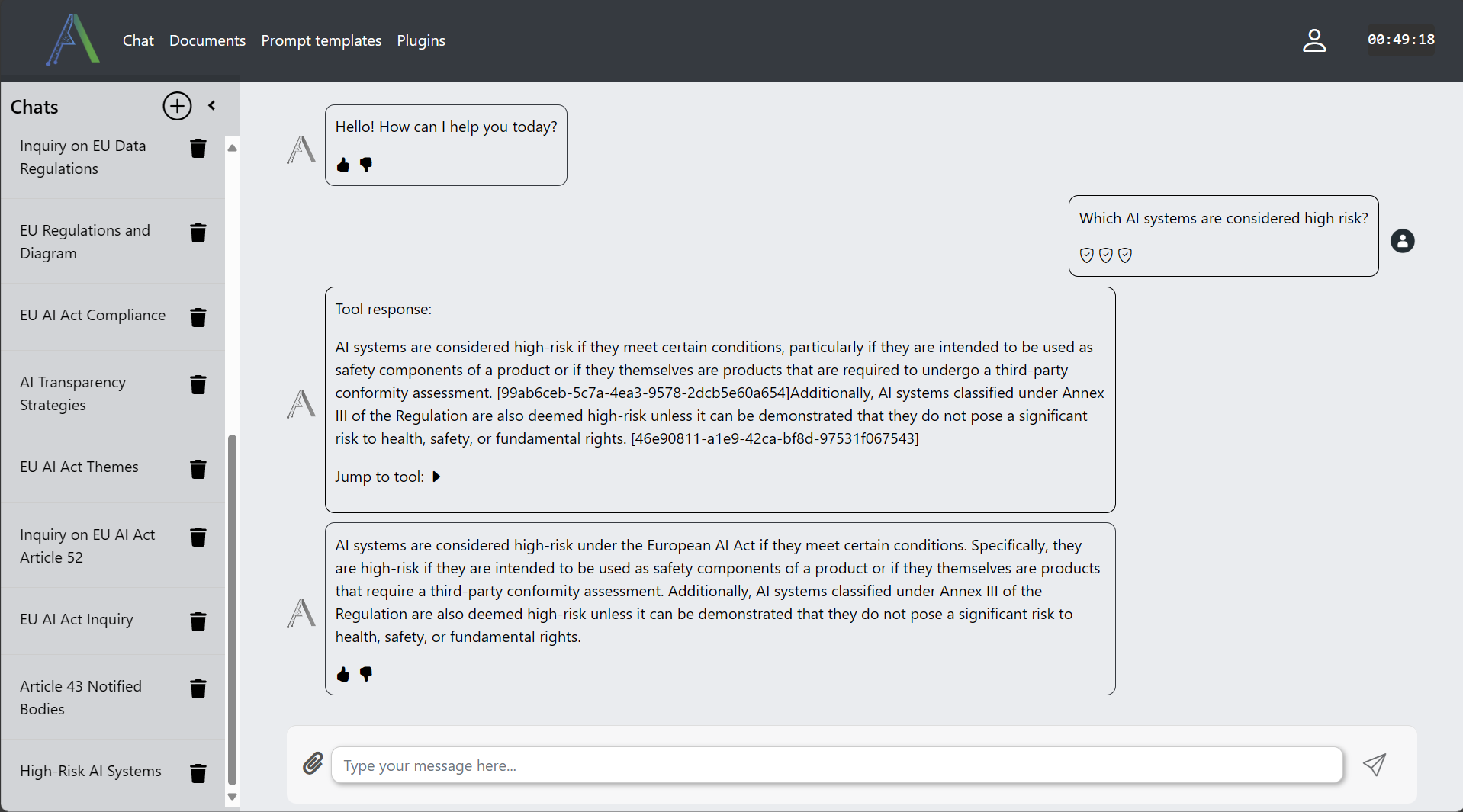}
    \caption{Simple question/answer pair from the interface of our developed chatbot}
    \label{fig:example-1}
\end{figure}
\subsubsection{Evaluation}

The operation of a naive RAG is based on the semantic similarity between the user query and the information extracted from the processed documents, therefore, it is quite effective when querying specific expressions within a relatively simple context.\looseness=-1

Manual evaluation of answer authenticity: To assess the reliability of the chatbot’s responses, we plan to validate its outputs against labeled ground-truth answers. One approach involves developing a structured questionnaire based on the EU AI Act. This questionnaire will help determine whether a given use case falls under the AI Act’s definition of AI, assess its risk classification (e.g., high-risk, low-risk), and identify the applicable regulatory requirements. Our legal experts will provide authoritative answers to this questionnaire based on the AI Act. We will then compare the chatbot’s responses with those provided by the legal experts to measure accuracy and alignment with the official regulatory framework.\looseness=-1

Industry use-case-based evaluation: Another evaluation approach involves collaboration with AI developers from the industry who have built AI-powered applications~\cite{almahasneh2024uncovering}. These developers will provide detailed descriptions of their use cases, which will be reviewed by our legal experts. The experts will identify the most relevant regulatory documents that need to be considered for each use case. The chatbot’s recommendations will then be compared against the expert-identified regulations to assess its effectiveness in real-world scenarios.\looseness=-1

The example shown in the Fig.~\ref{fig:example-1} clearly demonstrates that the naive RAG successfully identifies semantic similarities. In response to the simple conceptual question "Which AI systems are considered high risk?", it correctly identified the corresponding definitional section of the AI Act, accurately interpreted both the user query and the text segment containing the answer, and then generated the response.\looseness=-1
While this approach can yield satisfactory results in certain cases, it may not be suitable for more complex scenarios~\cite{microsoft_graphRAG_post}, i.e.:
\begin{itemize}
    \item When answering a question requires uncovering the relationships between disparate pieces of information based on their shared attributes in order to provide new, synthesized insights.
    \item When a holistic understanding of summarized semantic concepts in large datasets or documents is required.
\end{itemize}

\subsection{Graph RAG}
In contrast to the naive RAG, the Graph RAG does not search for semantic similarities in vector databases. Instead, it constructs knowledge graphs from related and unrelated data and their relationships from the source documents, which could enable a deeper level of interpretation of the source documents and with more complex questions, such as:
\begin{itemize}
    \item List the 8 areas included in the annex referenced in the definition of high-risk AI systems.
    \item Summarize the 5 most important topics of the AI Act!
\end{itemize}
Where semantic matching is not feasible, significant advancements can be achieved in answer generation~\cite{edge2024localglobalgraphrag}.\looseness=-1
While Naive RAG retrieves information chunks based on keyword similarity, which can lead to missing essential information chunks, the knowledge graph structure used by Graph RAG retrieves information not just based on semantic similarity but also on relationships between entities, ensuring more relevant and contextually rich retrieval.
Naive RAG processes documents as a collection of independent chunks, which may result in fragmented and less coherent responses. In contrast, Graph RAG enables reasoning over connected nodes, helping the model understand how different pieces of information relate, leading to more coherent and accurate answers. This capability also enables multi-hop retrieval, meaning the system can follow connections between related nodes, allowing it to access information that leads to a more insightful response—information that a traditional RAG model might have overlooked.
Naive RAG provides answers without offering much insight into why a particular piece of text was retrieved. In contrast, Graph RAG can trace responses back to structured knowledge, ensuring greater transparency and explainability in the answers.\looseness=-1

\begin{figure}[!htb] 
    \centering
    \includegraphics[width=0.5\textwidth]{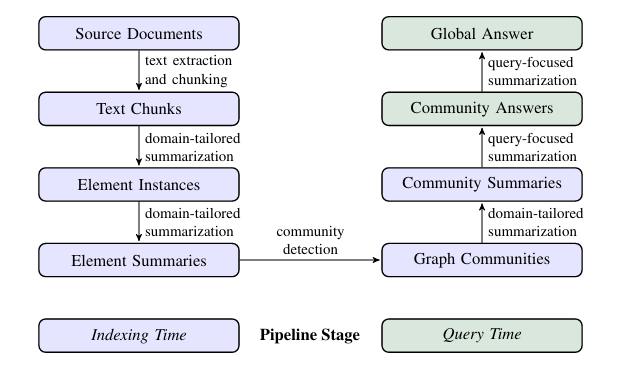}
    \caption{Graph-based RAG~\cite{edge2024localglobalgraphrag}}
    \label{fig:graph_rag_indexing}
\end{figure}
\begin{figure*}[!t] 
            \centering
            \includegraphics[width=0.75\textwidth]{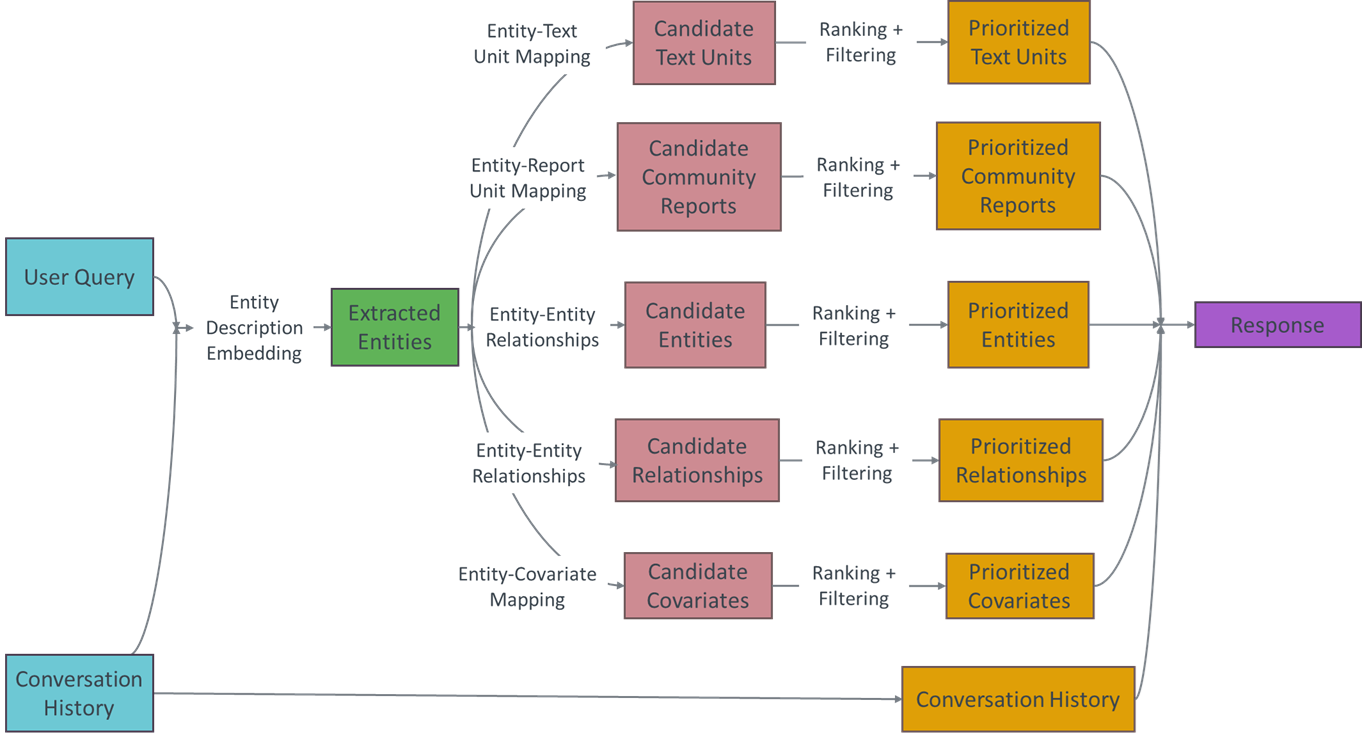}
            \caption{Local-search~\cite{Microsoft_GraphRAG_Local}}
            \label{fig:local_search}
\end{figure*}
\subsubsection{Indexing}
According to the indexing pipeline of Graph RAG model, as Fig.~\ref{fig:graph_rag_indexing} depicts, the processed documents must first be divided into chunks of a specified token size to ensure they are of manageable dimensions. This is necessary because, in the next step, an LLM is used to identify the entities, their relationships, and claims present in the source document required to construct the knowledge graphs.\looseness=-1

First, all entities in the text should be identified, including their name, type, and description. Following this, all relationships between clearly related entities should be identified, including the source and target entities and a description of their relationship. Next, claims should be extracted and linked to the detected entities, such as the subject, object, type, description, source text span, start- and end-dates. This process may require multiple iterations of LLM processing, and the definition of extraction prompts should take into account domain-specific features to ensure higher-quality results.\looseness=-1

In the second step, the instance-level summaries created must be transformed into a single block of text for each graph element (i.e., entity node, relationship edge, and claim-covariant). This requires an additional summarization round using the LLM for the corresponding instance groups.\looseness=-1

The resulting index can be modeled as a homogeneous, undirected, weighted graph, where the nodes represent entities, the edges represent the relationships between them, and the weights on the edges denote the count of these relationships. Given such a graph, a range of community detection algorithms may be employed to partition the graph into communities of nodes with stronger interconnections within the community than with nodes outside of it~\cite{edge2024localglobalgraphrag}. 

As the final step of the indexing phase, summaries must be created for the communities formed in the previous step, starting from the bottom of the hierarchy as follows:

\begin{itemize} 
    \item Leaf-level communities: The summaries of connected elements should be ordered in descending order on the basis of the degree of the source and target nodes. Then, iteratively, the context window is expanded by adding descriptions of the source, target, connected covariants, and edge. 
    \item Higher-level communities: If the summary of each element fits within the context window, the same procedure as the previous case should be applied. If not, the lower-level communities should be sorted in descending order based on the sum of the token counts of their element summaries and replace the summaries of the community elements (longer) with the community summary (smaller)~\cite{edge2024localglobalgraphrag}. 
\end{itemize}

The summaries of the communities are standalone and contain comprehensive semantic information of all the content of the elements lower in the hierarchy. If the user query asks for more detailed information, the relevant information can be found following the connection to the element to a deeper level of the hierarchy.

\subsubsection{Retrieval-Augmented Generation}
Once the knowledge graph has been constructed, graph RAG ensures inquiry in two ways, depending on the type of information requested by the user query. If it concerns the hollistic understanding of the whole data corpus, then global search in Fig.~\ref{fig:global_search} should be applied. If the user query is focused on interpreting specific entities and the relationships between them, then local search in Fig.~\ref{fig:local_search} should be used.

\begin{itemize}
    \item Global Search: the community summaries on a certain level should be random shuffled and divided into smaller chunks (to precede information loss), then intermediate answers are generated based on each chunk and these answers are rated by helpfulness between 0-100. In the final step these answers are sorted in descending order and injected into the context window until the token limit is reached to generate the final answer.
        \begin{figure*}[!t] 
            \centering
            \includegraphics[width=0.75\textwidth]{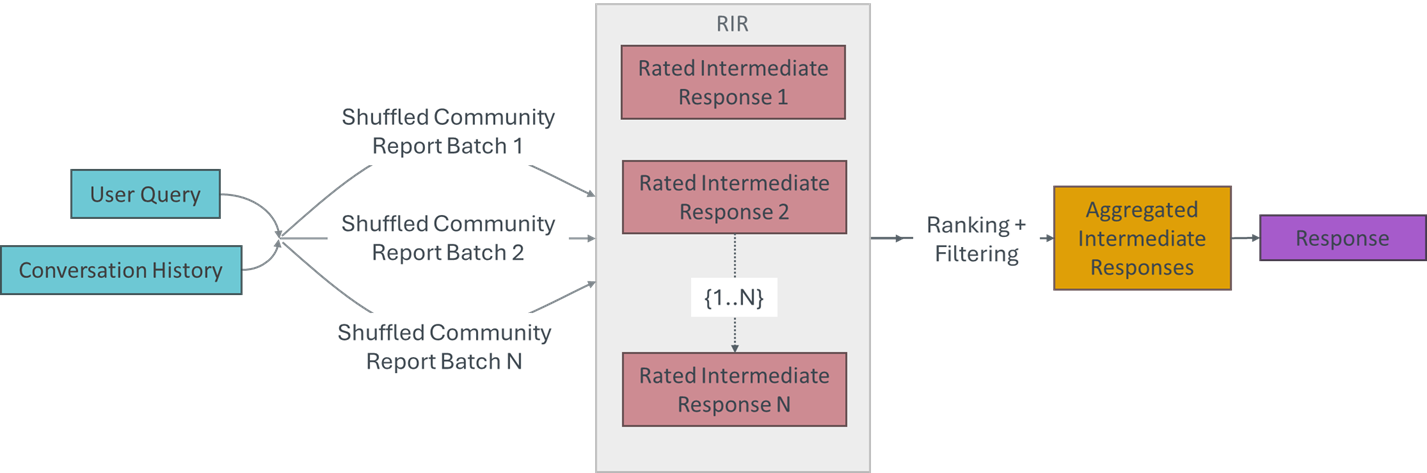}
            \caption{Global-search~\cite{Microsoft_GraphRAG_Global}}
            \label{fig:global_search}
        \end{figure*}

    \item Local Search: firstly entities are identified from the knowledge base based on semantic similarity to the user query. Subsequently, the entities, relationships, covariates, document segments, and community reports associated with the selected entities must be retrieved and prioritized. Based on the most relevant results, the final response can then be generated.
        
\end{itemize}

\subsubsection{Expectations}
Graph RAG enables substantial improvements in both comprehensiveness and diversity when compared to naive RAG models. Furthermore, in datasets or task-specific domains where global queries are more common, Graph RAG demonstrates a clear increase over naive RAG models, while achieving competitive results compared to other global models, all with reduced token consumption\cite{edge2024localglobalgraphrag}. We aim to further examine these aspects to generate increasingly accurate and effective responses within our self-assessment tool. 
All of this with the aim of assisting AI-related developers and engineers in their work, so that they are aware of the legal and ethical regulations that arise during development in relation to questions that arise, for example, from our industrial partner, who is supporting and strengthening human and manufacturing cycles through the development of AMRs (Autonomous Mobile Robots) and exoskeletons:

\begin{itemize}
    \item "What are the key safety standards and guidelines for AMRs with human detection systems to ensure safe interaction between robots and human workers in a factory setting?"
    \item "Are there ethical guidelines on the use of cameras in the workplace, specifically for human detection systems in AMRs, to prevent surveillance or misuse of data?"
\end{itemize}

\section{Future Direction and Conclusion}
This paper introduced an AI-driven self-assessment chatbot designed to support AI developers and stakeholders in navigating the evolving regulatory landscape. The current instance of our self-assessment tool covers the EU AI Act in particular, as it is the first binding regulation in the EU. By leveraging a Retrieval-Augmented Generation (RAG) framework, the chatbot enhances compliance verification by retrieving and interpreting relevant legal documents, reducing complexity, and ensuring adherence to ethical and regulatory standards.

Our implementation of both Naive RAG and Graph RAG architectures demonstrates the potential for improving AI governance tools. While the Naive RAG provides effective semantic matching for straightforward inquiries, the Graph RAG introduces a structured knowledge representation that facilitates deeper regulatory interpretation and more complex query resolution. The chatbot's ability to integrate proprietary and public regulations further enhances its value for organizations managing AI compliance.

Looking ahead, we anticipate three key developments in our work. 1) Regulatory Coverage: As regulatory landscapes evolve, we aim to expand our chatbot’s scope to encompass a wider range of regulations. This will not only ensure that user queries are addressed in compliance with binding regulations but also highlight the most critical documents users should be aware of. By integrating a more comprehensive set of legal frameworks, our chatbot will provide more effective guidance on compliance requirements. 2) Checklists as a complementary input: We are developing legal and ethical checklists based on existing regulations and guidelines. These checklists will be presented to users at the beginning of their interaction with the self-assessment tool. Users' responses to these checklists will serve as a complement to the prompts they submit to the chatbot, enhancing the chatbot’s ability to provide precise and contextually relevant guidance. This approach ensures that the tool delivers more comprehensive and tailored support, helping users navigate complex regulatory environments with greater accuracy and confidence. 3) Expand our implementation of the self-assessment tool by integrating it with Graph RAG, enabling precise answers to more complex prompts.

\bibliographystyle{IEEEtran}
\bibliography{Main}
\end{document}